\documentclass[12pt]{article}
\usepackage[english,german,french,polish]{babel}
\usepackage[T1]{fontenc}
\usepackage{amsfonts}

\begin{document}

\baselineskip 0.75cm
\topmargin -0.6in
\oddsidemargin -0.1in

\let\ni=\noindent

\renewcommand{\thefootnote}{\fnsymbol{footnote}}

\newcommand{\SM}{Standard Model }

\pagestyle {plain}

\setcounter{page}{1}

\pagestyle{empty}

~~~

\begin{flushright}
\end{flushright}

\vspace{0.3cm}

{\large\centerline{\bf Predicting the tauon mass}}

\vspace{0.5cm}

{\centerline {\sc Wojciech Kr\'{o}likowski}}

\vspace{0.3cm}

{\centerline {\it Institute of Theoretical Physics, University of Warsaw }}

{\centerline {\it Ho\.{z}a 69,~~PL--00--681 Warszawa, ~Poland}}

\vspace{1.2cm}

{\centerline{\bf Abstract}}

\vspace{0.3cm}

The recent experimental estimate for $\tau$ lepton mass comes significantly near to a theoretical value 
proposed by us in 1992. We recall our argumentation supporting this proposal.

\vspace{0.6cm}

\ni PACS numbers: 12.15.Ff , 12.90.+b .

\vspace{0.8cm}

\ni December 2009

\vfill\eject

~~~
\pagestyle {plain}

\setcounter{page}{1}

\vspace{0.2cm}

The central value of recent experimental estimate for $\tau$ lepton mass  [1],

\vspace{-0.2cm}

\begin{equation} 
 m^{(2008)}_{\tau} = 1776.84 \pm 0.17\;{\rm MeV} \,,
\end{equation} 

\ni is considerably diminished in comparison with that of the former experimental estimation [2],  

\vspace{-0.2cm}

\begin{equation} 
 m^{(2006)}_{\tau} = 1776.99^{+0.29}_{-0.26}\;{\rm MeV} \;,
\end{equation} 

\ni while the errors are reduced a lot. As it happens, in 1992 we proposed for $m_\tau$ a theoretical value [3],

\vspace{-0.2cm}

\begin{equation}
m_\tau  = 1776.80\;{\rm MeV} \;,
\end{equation}

\ni very close to the actual estimate (1). We deduced such a value from the parameter-free mass sum rule [3]
 
\vspace{-0.2cm}

\begin{equation}
125 m_\tau = 6 \left(351m_\mu - 136 m_e\right) 
\end{equation}

\ni  resulting strictly from a mass formula conjectured in 1992 for charged leptons $ e^-, \mu^-, \tau^-$. In fact, the sum rule (4) implies the value (3), when the experimental figures for $m_e$ and $m_\mu$ are used as an only input (it gives $m_\tau = 1776.797$ MeV with $m_e = 0.5109989$ MeV and $m_\mu = 105.6584$ MeV [1,2]).{\footnote{ In 1981, Koide proposed for charged leptons the neatly looking nonlinear equation to describe their mass spectrum [4],
$$
m_e + m_\mu + m_\tau = \frac{2}{3}(\sqrt{m_e} + \sqrt{m_\mu} + \sqrt{m_\tau})^2,
$$
having --- as can be seen --- two solutions for $m_\tau$ in terms of $m_e$ and $m_\mu$,
$$
m_\tau = \left[2(\sqrt{m_e} + \sqrt{m_\mu} )\pm \sqrt{3(m_e + m_\mu)+12\sqrt{m_e m_\mu}} \right]^2 = \left\{\begin{array}{r} 1776.97\,{\rm MeV} \\ 3.31735\,{\rm MeV}  \end{array} \right.\,,
$$ 
where $m_e$ = 0.5109989 MeV and $m_\mu$ = 105.6584 MeV are experimental figures. The first solution agrees wonderfully with the central value of actual experimental estimate (1), though its small deviation from this experimental value is slightly larger than the deviation of our prediction (3) (for the former experimental estimation (2), the situation was opposite). The second solution gets no interpretation.}}

In the present note we would like to recall our mass formula for charged leptons and then, in Appendix, review some arguments supporting its form.

Our charged-lepton mass formula expresses three masses $m_e , m_\mu , m_\tau$ in terms of two free parameters $\mu$ and $\mu(\varepsilon -1)$, one running quantum number $N$ taking three values

\vspace{-0.1cm}

\begin{equation}
N = 1,3,5
\end{equation}

\ni and, finally, three generation-weighting factors

\vspace{-0.1cm}

\begin{equation} 
\rho_1 = \frac{1}{29} \;,\; \rho_3 = \frac{4}{29} \;,\; \rho_5 = \frac{24}{29} 
\end{equation}

\ni ($\sum_N\rho_N = 1$). With the notation

\vspace{-0.1cm}

\begin{equation}
m_1  = m_e\;,\;  m_3  = m_\mu\;,\;  m_5  = m_\tau
\end{equation}

\ni our formula reads [3]

\vspace{-0.2cm}

\begin{equation}
m_N  =  \rho_N \,\mu \left(N^2 + \frac{\varepsilon -1}{N^2} \right) \;\;\;\;(N = 1,3,5)
\end{equation}

\ni or, more explicitly,

\vspace{-0.1cm}

\begin{eqnarray}
m_e & = & \frac{\mu}{29} \,\varepsilon  \,, \nonumber \\
m_\mu & = & \frac{4\mu}{29}\,\frac{80 +\varepsilon}{9}  \,, \nonumber \\
m_\tau & = & \frac{24\mu}{29}\,\frac{624 + \varepsilon}{25} \,. 
\end{eqnarray}

It can be seen that, eliminating two free parameters $\mu$ and $\varepsilon$ from the mass formula (9), we obtain strictly the mass sum rule (4) expressing $m_\tau$ in terms of $m_e$ and  $m_\mu$. On the other hand, from the first and second Eq. (9) we can evaluate $ \mu$ and $\varepsilon$  also in terms of $m_e$ and  $m_\mu$,

\vspace{-0.1cm}

\begin{equation}
\mu = \frac{29 (9m_\mu - 4 m_e)}{320} \;\;,\;\;\; \varepsilon = \frac{320 m_e}{9 m_\mu - 4 m_e} \,,
\end{equation}

\ni and hence, numerically

\vspace{-0.1cm}

\begin{equation}
\mu = 85.9924\;{\rm MeV} \;\;\;,\;\;\; \varepsilon = 0.172329\,
\end{equation}

\ni with the experimental figures for $m_e$ and $m_\mu$ used as an only input. Then, our mass formula (8) or (9) is fully defined to determine $m_e$ and $m_\mu$ equal to their experimental figures and $m_\tau$ equal to its predicted value (3).

In Appendix, we present some arguments supporting the use of $N = 1,3,5$ as the adequate quantum number and the special choice of $\rho_N = 1/29, 4/29 , 24/29$ as the generation-weighting factors as well as the specific dependence of charged-lepton masses on $N^2$ and $1/N^2$. These arguments lead to our mass formula (8) or (9).

We believe that the discovery of explicit formulae for masses of fundamental fermions is still possible, as it was for the levels of hydrogen atom in Balmer's and Bohr's time, although now we have to operate with the involved quantum field theory or its extensions, rather than with the simple rudiments of quantum mechanics as in the old times. 


~~~
\vspace{0.3cm}

{\centerline {\bf Appendix}}

{\centerline {\it Three generations of fundamental fermions caused by an intrinsic Pauli principle,}}

{\centerline {\it and a charged-lepton mass formula}}

\vspace{0.3cm}

 Assume that fundamental fermions, such as \SM leptons and quarks, are pointlike objects in spacetime, but intrinsically composite structures in an algebraic sense expressed by the multicomponent form of their local one-particle wave function 

$$
\psi_{\alpha_1 \alpha_2... \alpha_N}^{(N)}(x)\;\;\;\;(N = 1,3,5,\ldots)\,.
\eqno{\rm (A1)}
$$

\ni Here, $\alpha_i = 1,2,3,4 \;(i=1,2,\ldots,N)$ are Dirac bispinor indices in the chiral representation, where 
commuting $4N\times 4N$ Dirac matrices $\gamma^{5}_i$ and $\sigma^{3}_i \;(i=1,2,\ldots,N)$ are diagonal. Here, $ \gamma_j^5 = i \gamma_j^0 \gamma_j^1 \gamma_j^2 \gamma_j^3$ and $\sigma_j^3 = \gamma_j^5 \gamma_j^0 \gamma_j^3$. If the basic $4N\times 4N$ Dirac matrices $\gamma^{\mu}_i\; (i = 1,2,\ldots,N)$ are elements of Clifford algebra $\left\{ \gamma^{\mu}_i\,,\,\gamma^{\nu}_j  \right\} = 2 g^{\mu \nu} \delta_{i j}\;\; (i,j = 1,2,\ldots,N)$, anticommuting for $i \neq j$, it turns out that the wave function (A1) can be assumed to satisfy in the free case the wave equation of generalized free Dirac form [5],

$$
\left( \Gamma^{(N) \mu}\, p_\mu -  M^{(N)}\right) \psi^{(N)}(x) = 0 \;\;\;\;(N = 1,3,5,\ldots)\,,
\eqno{\rm (A2)}
$$

\ni where $\psi^{(N)}(x) \equiv \left( \psi_{\alpha_1 \alpha_2... \alpha_N}^{(N)}(x)\right)$, while 

\vspace{-0,1cm}
 
$$
\Gamma^{(N) \mu} \equiv \frac{1}{\sqrt{N}} \left(\gamma^{\mu}_1  + \gamma^{\mu}_{2} + \ldots + 
\gamma^{\mu }_{N} \right)   \;\;(N = 1,3,5,\ldots)
\eqno{\rm (A3)}
$$

\ni are elements of Dirac algebra $\{ \Gamma^{(N) \mu}\, ,  \,\Gamma^{(N) \nu}\} = 2g^{\mu \nu}$ following from Dirac's square root procedure $\sqrt{p^2}\rightarrow \Gamma^{(N) \mu}\, p_\mu$. Here, all possible \SM  $ SU(3)\times SU(2)\times U(1) $ labels are suppressed. 

Because of the form (A3) of $\Gamma^{(N) \mu}$ it will be convenient to pass in Eq. (A2) from the individual Dirac matrices $\gamma^\mu_i \;\; (i = 1,2,\ldots,N)$ to their Jacobi-type combinations

\begin{eqnarray*}
\Gamma^{(N) \mu}_1 & \equiv & \frac{1}{\sqrt{N}} \left(\gamma^\mu_1 + \gamma^\mu_2 + \ldots + \gamma^\mu_N \right) \equiv \Gamma^{(N)\,\mu} \;, \\ 
\Gamma^{(N) \mu}_i & \equiv & \frac{1} {\sqrt{i(i - 1)}} \left[ \gamma^\mu_1 + \gamma^\mu_2 + \ldots + \gamma^\mu_{i-1} - (i - 1) \gamma^{\mu}_i \right] \\ 
 & & \;\;\;\;\;\;\;\;\;\;\;\;\;\;\;\;(i = 2,3,\ldots, N) \,,
\end{eqnarray*} 
\vspace{-2.37cm}

\begin{flushright}
(A\,4)
\end{flushright}

\ni satisfying the Clifford algebra $\left\{ \Gamma^{(N) \mu}_i\, ,  \,\Gamma^{(N) \nu}_j \right\} = 2g^{\mu \nu} \delta_{i j} \;\;\; (i,j = 1,2,\ldots,N)$ isomorphic with the previous Clifford algebra. The combinations (A4) may be called "$\!$\,centre-of-mass"\, and "\,$\!$relative"\, Jacobi-type Dirac matrices, respectively. In the new chiral representation, where commuting Jacobi-type Dirac matrices $\Gamma^{(N) 5}_i$ and $\Sigma^{(N) 3}_i \;(i = 1,2,\ldots, N)$, analogical to $\gamma^5_i$ and $\sigma^3_i$, are diagonal and $\alpha_i = 1,2,3,4 \;(i=1,2,\ldots,N)$ denote new Dirac's bispinor indices, the free generalized Dirac equation (A 2) can be reduced to the form [5]

$$
\left( \gamma^{\mu}\, p_\mu -  M^{(N)}\right)_{\alpha_1 \beta_1} \psi^{(N)}_{\beta_1 \alpha_2... \alpha_N}(x) = 0 \;\;\;\;(N = 1,3,5,\ldots)
\eqno{\rm (A 5)}
$$

\ni with $\psi^{(N)}(x) \!\!\equiv  \!\!\left( \!\psi_{\alpha_1 \alpha_2... \alpha_N}^{(\!N\!)}(x)\!\right)$. Here, $\!\gamma^\mu\!$ are the ordinary $4\!\times\! 4$ Dirac matrices and $\Gamma_1^{(N) \mu}\!\! = \gamma^\mu\otimes {\bf 1}\otimes\dots\otimes {\bf 1}$. Note that in Eq. (A\,5) $\alpha_1$ is the "\,$\!$centre-of-mass"\, Dirac bispinor index, while $\alpha_2,...,\alpha_N$ present "\,$\!$relative"\, Dirac bispinor indices (see the definitions (A 4)).

We can introduce to the free wave equation (A 5) the \SM gauge interactions applying the minimal substitution $ p_\mu \rightarrow p_\mu - g A_\mu(x)$, where $A_\mu(x)$ involves the familiar weak-isospin and color matrices, the weak hypercharge dependence as well as the ordinary chiral $4\times 4$  Dirac  matrix $\gamma^5$ (in general, $\Gamma^{(N) 5}_1$). Then, $g\, \gamma^\mu A_\mu(x)$ (in general, $g \,\Gamma_1^{(N) \mu} A_\mu(x)$) becomes the \SM gauge coupling in the generalized Dirac equation which can be reduced to the form [5] 

$$
\left\{ \gamma^\mu \left[p_\mu - g A_\mu(x)\right] -  M^{(N)}\right\}_{\alpha_1 \beta_1} \psi^{(N)}_{\beta_1 \alpha_2... \alpha_N}(x) = 0 \;\;(N = 1,3,5,\ldots)\,.
\eqno{\rm (A 6)}
$$

\ni This is a coupling to the "$\!$\,centre-of-mass"\, Dirac bispinor index $\alpha_1$. It is plausible to assume that also  all other possible interactions, including the familiar gravity and hypothetic hidden-sector interactions, if effective, are coupled to the fundamental fermions {\it via} their "$\!$\,centre-of-mass"\,  degrees of freedom $\alpha_1$ and $p_\mu$. Hence, the conclusion follows that the "$\!$\,centre-of-mass"\, bispinor index $\alpha_1$ is physically distinguished from all "\,$\!$relative"\, bispinor indices $\alpha_2,\ldots, \alpha_N$ which, being uncoupled, are mutually undistinguishable. Thus, it is very natural to conjecture that all "\,$\!$relative"\, bispinor indices $\alpha_2,\ldots, \alpha_N$, treated as intrinsic physical objects ("\,$\!$intrinsic partons"), obey the Fermi statistics along with the {\it intrinsic Pauli principle} [5] requiring full antisymmetry of one-particle wave function $\psi^{(N)}_{\alpha_1 \alpha_2... \alpha_N}(x)$ with respect to all $\alpha_2,\ldots, \alpha_N$ indices. In consequence, $N$ becomes restricted to its three values $N =1,3,5$ only.

It is exciting to observe that the above simple conjecture of full antisymmetry of the wave function (A 1) with respect to all "\,$\!$relative"\, bispinor indices $\alpha_2,\ldots, \alpha_N$ implies the existence in Nature of exactly {\it three generations} $N =1,3,5$ of fundamental fermions with spin $1/2$ (including \SM leptons and quarks [5] as well as sterile fermions with spin 1/2, we have called {\it sterinos}, candidates for the stuff the cold dark matter is made of [6]). The one-particle wave function of these fermions in three generations can be written down as the following Dirac bispinors [3]:

\begin{eqnarray*}
& & \psi^{(1)}_{\alpha_1}(x) \;\,,  \\
& & \psi^{(3)}_{\alpha_1}(x) = \frac{1}{4}\left(C^{-1} \gamma^5 \right)_ {\alpha_2 \alpha_3} \psi^{(3)}_{\alpha_1 \alpha_2 \alpha_3}(x) =  \psi^{(3)}_{\alpha_1 1 2}(x) =  \psi^{(3)}_{\alpha_1 3 4}(x)\;\,, \\
& & \psi^{(5)}_{\alpha_1}(x) = \frac{1}{24}\varepsilon_{\alpha_2 \alpha_3 \alpha_4 \alpha_5} \psi^{(5)}_{\alpha_1 \alpha_2 \alpha_3 \alpha_4 \alpha_5}(x) = \psi^{(5)}_{\alpha_1 1 2 3 4}(x)  \;\,, 
\end{eqnarray*} 

\vspace{-1.58cm}

\begin{flushright}
(A 7)
\end{flushright}

\ni where, in addition,

$$
\psi^{(3)}_{\alpha_1 1 3}(x) = 0 = \psi^{(3)}_{\alpha_1 2 4}(x)\;\;,\;\;\psi^{(3)}_{\alpha_1 1 4}(x) = 0 = \psi^{(3)}_{\alpha_1 2 3}(x)\;.
\eqno{\rm (A 8)}
$$

\ni In Eqs. (A\,7) and (A\,8), the probability interpretation and relativity of quantum theory are applied [3]. Here, the \SM $ SU(3)\times SU(1)\times U(1) $ labels --- accompanying the "\,$\!$centre-of-mass"\, bispinor index $\alpha_1$ in the case of leptons and quarks --- are suppressed.  

Note that the total "\,$\!$relative"\,chirality $\Gamma^{(N) 5}_2\!\ldots \!\Gamma^{(N) 5}_N\;\,(N \!=\! 3,5)$ commutes with the Hamiltonian following from the generalized Dirac equation [5] (see Eq. (A\,6))

$$
\left\{\Gamma^{(N) \mu}_1 \left[p_\mu - g A_\mu(x)\right] -  M^{(N)}\right\}\psi^{(N)}(x) = 0 \;\;\;\;\;(N = 1,3,5)\,, \eqno{\rm (A 9)}
$$

\ni so it is a constant of motion (note that $\Gamma^{(N) 5}_1 $ and $\Gamma^{(N) 5}_2,\ldots,\Gamma^{(N) 5}_N$ anticommutes and commute with $\Gamma^{(N) \mu}_1$, respectively, while all $\Gamma^{(N) 5}_i \;(i=1,2,\ldots,N)$ commute mutually and also with $\Gamma^{(N) 0}_1\Gamma^{(N) \mu}_1$). This allows us to impose on the wave function $\psi^{(N)}(x) = \left(\psi^{(N)}_{\alpha_1 \alpha_2 \ldots \alpha_N} (x)  \right)$ the constraint $\Gamma^{(N) 5}_2\ldots \Gamma^{(N) 5}_N \psi^{(N)}(x)  = \psi^{(N)}(x) \;(N = 3,5)$, in consistency with the wave equation (A 9). However, there is no need to do it, because such a constraint is already fulfilled by the wave functions for $N = 3,5$ given in Eqs. (A 7) and, in addition, is consistent with Eqs. (A 8) for $N = 3$. The last observation excludes from the generation $N = 3$ the "\,$\!$relative"\, spins 1 and 0 with "\,$\!$relative"\, chirality $-1$ as $ 1 = {(+\uparrow)} , 2 = (+\downarrow) , 3 = (-\uparrow) , 4 = (-\downarrow)$, leaving in both generations $N =3,5$ only the "\,$\!$relative"\, spin 0 with "\,$\!$relative"\,  chirality +1 (the "\,$\!$relative"\, spin 1 with "\,$\!$relative"\, chirality +1 is directly excluded by the intrinsic Pauli principle giving $\psi^{(3)}_{\alpha_1 11}(x) = \psi^{(3)}_{\alpha_1 22}(x) = \psi^{(3)}_{\alpha_1 33}(x) = 0$). Notice that this constraint is natural, since it guarantees the same total chirality for all three generations $N=1,3,5$: $\Gamma^{(N) 5}_1\Gamma^{(N) 5}_2\ldots \Gamma^{(N) 5}_N \psi^{(N)}(x) = \Gamma^{(N) 5}_1 \psi^{(N)}(x)  = \gamma^5 \psi^{(N)}(x) $, where $\gamma^5 = \left(\gamma^5_{\alpha_1 \beta_1}\right)$. 

From Eqs. (A7) it follows that

$$
\psi_{\alpha_1 \alpha_2... \alpha_N}^{(N)*}(x) \psi_{\alpha_1 \alpha_2... \alpha_N}^{(N)}(x) = 29 \rho_N \psi_{\alpha_1}^{(N)*}(x) \psi_{\alpha_1}^{(N)}(x)\;\;\;\;(N = 1,3,5)\,,
\eqno{\rm (A 10)}
$$

\ni  where

$$
\rho_1 = \frac{1}{29} \;,\; \rho_3 = \frac{4}{29} \;,\; \rho_5 = \frac{24}{29} 
\eqno{\rm (A 11)}
$$

\ni  ($\sum_N\rho_N = 1$) are the generation-weighting factors.

Now, restricting ourselves to charged leptons of three generations, $e^-, \mu^-, \tau^-$, as the simplest \SM fermions, we will look for their mass formula in the form [5]

$$
m_N = \rho_N h_N \;\; \;\; (N = 1,3,5)  \,,
\eqno{\rm (A 12)}
$$

\ni where $m_e = m_1 , m_\mu = m_3 , m_\tau = m_5$ and $\rho_N$ denote three generation-weighting factors (A~11). We will put 

$$
h_N = \mu u_N + \mu(\varepsilon -1) v_N \;\; \;\; (N=1,3,5)  
\eqno{\rm (A 13)}
$$

\ni in terms of two massdimensional parameters $\mu$ and $\mu(\varepsilon-1)$ which combine additively two  "\,$\!$intrinsic interactions"\, $\mu u_N$ and $\mu(\varepsilon-1) v_N$ "\,$\!$within"\, charged leptons.

In order to build up these "\,$\!$intrinsic interactions"\, we have to our disposal only three structural numbers $N$ (three other structural numbers $\rho_N$, appearing in our approach, are already used in the formula (A~12)). We will conjecture that these formal interactions are: 

\begin{description}
\item{(i)} {~"\,$\!$intrinsic two-body interaction"\, between all "\,$\!$intrinsic partons"\, $i=1,2,\ldots,N$, treated on an equal footing:

\vspace{-0.2cm}

$$
\mu\, u_N = \mu \sum_{i,j=1}^N 1 = \mu N^2 = \mu N + \mu N(N-1) \;\;\;\;\;(N = 1,3,5) \,, 
\eqno{\rm (A 14)}
$$

\ni where the first term $\mu N$ on the rhs describes the sum of "\,$\!$intrinsic self-interactions"\, of all  "\,$\!$intrinsic partons", while the second term $\mu N(N-1)$ presents the sum of their  "\,$\!$intrinsic mutual interactions",  }

\item{(ii)} {~a correction to the  "\,$\!$intrinsic self-interaction"~of the "\,$\!$centre-of-mass intrinsic parton" 
\, $i=1$ distinguished from all "\,$\!$relative intrinsic partons"\, $\!i=2,\!\ldots\!,N$ which, in turn, are undistinguishable from each other:

\vspace{-0.2cm}

$$
\mu(\varepsilon -1) v_N = \mu(\varepsilon-1)\! {\left(P^{(N)}_{i=1}\!\right)}^{\!2}\! = \mu(\varepsilon-1) \frac{1} {N^2}  \;\;(N = 1,3,5)\,,
\eqno{\rm (A 15)}
$$

\vspace{-0.1cm}

\ni where $P^{(N)}_{i=1} = [N!/(N-1)!]^{-1} = 1/N$ is the probability of finding such a distinguished "\,$\!$intrinsic parton"~among $\!N\!$ "$\,\!$intrinsic partons"~ of which $N\!-1\!$ are undistinguishable.}
\end{description}

The forms (A\,12) and (A\,13) together with the conjectures (A\,14) and (A\,15) provide for charged leptons $e^-, \mu^-, \tau^-$ the mass formula (8) or (9):

\vspace{-0.2cm}

$$
m_N = \mu \,\rho_N \left(N^2 + \frac{\varepsilon - 1}{N^2}\right) \;\; \;\; (N = 1,3,5)  \,,
\eqno{\rm (A 16)}
$$

\ni or, more explicitly,

\vspace{-0.3cm}

\begin{eqnarray*}
m_e & = & \frac{\mu}{29} \,\varepsilon  \,, \nonumber \\
m_\mu & = & \frac{4\mu}{29}\,\frac{80 +\varepsilon}{9}  \,, \nonumber \\
m_\tau & = & \frac{24\mu}{29}\,\frac{624 + \varepsilon}{25} 
\end{eqnarray*}

\vspace{-0.1cm}

\vspace{-1.56cm}

\begin{flushright}
(A 17)
\end{flushright}

\ni with $m_e = m_1 , m_\mu = m_3 , m_\tau = m_5$ and $\rho_1 = 1/29 , \rho_3= 4/29 , \rho_5 = 24/29$.

Some extensions of the charged-lepton mass formula (A\,16) or (A\,17) were discussed in the two last Refs. [5]. They seem to be redundant in the case of charged leptons.

The intrinsic structure of fundamental fermions presented in this Appendix is an analogy of Dirac's construction of spin 1/2 that can be considered as an act of algebraic abs\-traction from the spatial notion of angular momentum, rather than a result of a point\-like approximation for a spatially extended rotating top. Similarly, our intrinsic construction of fundamental fermions can be considered as arising through an act of algebraic abstraction from the spatial notion of composite systems, rather than resulting from a pointlike approximation for spatially extended composite states. 

However, from the purely phenomenological point of view, the Dirac bispinor indices $\alpha_1, \alpha_2, \ldots, \alpha_N$ in the wave function (A\,1) may be only the summit of an iceberg of some, yet invisible, spatial partons bound in really composite systems, mainly in $S$ states. But, if the spatial partons were some conventional Dirac fermions, the Dirac matrices $\gamma^\mu_i$ and $\gamma^\nu_j$ should commute for $i \neq j$ (not anticommute as in our case; in both cases, all $\gamma^5_j \equiv i\gamma^0_j\gamma^1_j\gamma^2_j\gamma^3_j$ and $\sigma^3_j \equiv \gamma^5_j \gamma^0_j \gamma^3_j$  $(j=1,2,\ldots,N)$ commute). This would spoil our intrinsic construction based on the Clifford algebra for $\gamma^\mu_i\;\; (i = 1,2,,\ldots N)$. In this case, the wave function (A\,1) would not satisfy the generalized Dirac equation, though such a equation should follow from Dirac's square-root procedure hopefully accepted for fundamental fermions in the limit of local one-particle wave function. This argument seems to support our choice of intrinsically composite fundamental fermions described in this Appendix.

\vfill\eject

~~~~
\vspace{0.5cm}

{\centerline{\bf References}}

\vspace{0.5cm}

{\everypar={\hangindent=0.6truecm}
\parindent=0pt\frenchspacing

{\everypar={\hangindent=0.6truecm}
\parindent=0pt\frenchspacing

~[1]~C.~Amsler {\it et al.} (Particle Data Group), {\it Review of Particle Physics, Phys. Lett.} {\bf B 667}, 1 (2008).

\vspace{0.2cm}

~[2]~W.M.~Yao {\it et al.} ( Particle Data Group), {\it Review of Particle Physics, J. Phys}, {\bf G 33}, 1 (2006).

\vspace{0.2cm}

~[3]~W. Kr\'{o}likowski, {\it Acta Phys. Pol.} {\bf B 23}, 933 (1992); and references therein. In that paper, two options of the mass sum rule, $125 m_\tau = 6(351 m_\mu \mp 136 m_e)$, predicting $m_\tau =$ ( 1776.80 or 1783.47) MeV, respectively, were discussed (the first was {\it a priori} preferred, as it turns out now, correctly).

\vspace{0.2cm}

~[4]~For a more recent discussion {\it cf.} Y.~Koide, {\tt hep--ph/0506247}; and references therein.

\vspace{0.2cm}

~[5]~W. Kr\'{o}likowski, {\it Acta Phys. Polon.} {\bf B 21}, 871 (1990);  {\it Phys. Rev.} {\bf D 45}, 3222 (1992); {\it Acta Phys. Polon.} {\bf B 33}, 2559 (2002); {\tt hep--ph/0504256}; {\tt hep--ph/0604148}; {\it Acta Phys. Polon.} {\bf B 37}, 2601 (2006); {\it Acta Phys. Polon.} {\bf B 38}, 3133 (2007). 

\vspace{0.2cm}

~[6]~W. Kr\'{o}likowski, arXiv: 0811.3844 [{\tt hep-ph}]; and references therein. 

\vspace{0.2cm} 

\vfill\eject

\end{document}